\documentclass[12pt]{iopart}
\usepackage{graphicx,graphics}
\usepackage{iopams}
\usepackage{harvard}

\begin{document}
\bibliographystyle{jphysicsB}
\title[Electron collisions with C$_2^-$] {Low and intermediate energy
electron collisions with the C$_2^-$ molecular anion}
 
\author{Gabriela Halmov{\'a}$^{\dag}$ , J.D. Gorfinkiel$^\ast$  and Jonathan Tennyson$^{\dag}$ }
\address{$^{\dag}$ Department of Physics and Astronomy,
        University College London, Gower St.,
       London WC1E 6BT, UK\\
$^\ast$ Department of Physics and Astronomy, The Open University,
 Walton Hall, MK7 6AA Milton Keynes, UK}
\begin{abstract}
Calculations are presented which use the molecular R-matrix with
pseudo-states (MRMPS) method to treat electron impact electron
detachment and electronic excitation of the carbon dimer
anion. Resonances are found above the ionisation threshold of C$_2^-$
with $^1\Sigma^+_g$, $^1\Pi_g$ and $^3\Pi_g$ symmetry. These
are shape resonances trapped by the effect of an attractive
polarisation potential competing with a repulsive Coulomb interaction.  The
$\Pi_g$ resonances are found to give structure in the detachment cross
section similar to that observed experimentally.  Both excitation and
detachment cross sections are found to be dominated by large impact
parameter collisions whose contribution is modelled using the Born
approximation.
\end{abstract}
 
\section{Introduction}
 
The carbon dimer anion is unusual among small anions in having bound electronically excited states. In addition electron scattering experiments performed
in storage rings showed that C$_2^-$ was one of a number
of diatomic anions which, at least temporarily, are able
to bind an additional electron leading to pronounced resonance structures
in their measured cross sections  \cite{ahk96,pdj98,pdj99,abb01,cpf05}.
Similar results have also been found for systems containing the NO$_2^-$
anion \cite{ahk01,sbe05}.
Given that the electron collisions occur against the background of a
strongly repulsive Coulomb interaction, the occurrence of such resonance
structures, which generally lie above the threshold for electron
impact detachment, was unanticipated.

Theoretically the treatment of electron collisions with these diatomic
anions is complicated by the need to treat the region immediately above the
electron detachment threshold. Such intermediate energy calculations
are difficult because of the presence of two continuum electrons.
Previous theoretical studies have used bound state methods either
unadapted \cite{pdj99} or with an absorbing potential
\cite{srm97,stm00} to study the continuum states of C$_2^{2-}$.

In a preliminary publication \cite{jt450} we reported calculations using
the molecular R-matrix with Pseudo-states (MRMPS) method
\cite{jt341,jt354} which found a number of low-lying resonances. In
this work we report fully on these calculations, including details of the
models we tested to produce stable calculations and to
demonstrate that our results are robust. In addition we report results
for electron impact excitation of C$_2^-$.

\section{Method}

The \emph{R}-matrix method is based on dividing coordinate space into
two regions using a spherical boundary of radius $a$ 
centred on the centre-of-mass of 
the target molecule \cite{bb93,jt351}. 
The radius of the boundary is chosen so that the 
inner region contains all the electronic cloud of the target molecular 
states included in the calculation.

Inside the \emph{R}-matrix sphere it is necessary to consider all
short-range interactions between the \emph{N} target electrons and the 
scattering one, such as exchange and electron correlation. In the outer 
region these effects are negligible so the scattering electron can be 
described as moving in the long-range multipole potential of the target 
molecule.

The accuracy of this method is strongly dependent on the representation 
of the problem in the inner region \cite{jt189}. 
In standard, low-energy calculations, the wave function for an
\emph{N}+1 electrons system in the inner region is given by the expansion:
\begin{equation}
       \psi_k^{N+1} = \mathcal{A} \sum_{ij} a_{ijk} \Phi_i (\textbf{x}_1...
\textbf{x}_N) u_{ij} (\textbf{x}_{N+1}) + \sum_i b_{ik} \chi_i(\textbf{x}_1...\textbf{x}_{N+1})\:,\label{eq:rmat}
\end{equation}
where $k$ represents the $k^{\rm th}$ solution of the inner region
Hamiltonian, $\mathcal{A}$ is the antisymmetrisation operator,
$\textbf{x}_i$ are the spatial and spin coordinates of 
electron \emph{i}, \emph{u$_{ij}$} are continuum orbitals (COs) which
represent the scattering electron \cite{jt238}, \emph{a$_{ijk}$} and
\emph{b$_{ik}$} are variational coefficients, \emph{$\Phi_i$} is the
wave function of the $i^{\rm th}$ target state and
\emph{$\chi_i$} are \emph{L$^2$} functions constructed from the
target occupied and virtual molecular orbitals. These functions represent 
electron correlation and polarisation effects. In the first sum, 
the configuration state functions are constrained to 
give the correct (target) space and
spin symmetry for the first $N$-electrons as well as the correct total, $N+1$
electron space-spin symmetry. Doing this requires special consideration
of phase effects due to electron ordering in the wave function \cite{jt195}.

In the standard formulation of the R-matrix method all the solutions of
the inner region problem are required to construct the R-matrix on the
boundary \cite{bb93}. This presents a serious computational 
barrier for large calculations as diagonalising the entire Hamiltonian
may not be feasible. For some of the calculations discussed 
below we used the partitioned R-matrix method \cite{jt332}, which 
only requires the low-lying eigenvalues and which has been demonstrated to
give good results for the electron -- C$_2^-$ problem \cite{jt419}. 

Here we use the UK polyatomic \emph{R}-matrix code
\cite{jt204} rather than the specialised, Slater orbital-based diatomic code.
This is because the MRMPS method  described
below relies on the use Gaussian Type Orbitals (GTOs) and is therefore
only implemented in this code. The highest symmetry available in the
polyatomic code is D$_{2h}$ which is a subgroup of the true D$_{\infty
h}$ symmetry of C$_2^-$. All calculations presented here were performed in
D$_{2h}$ symmetry. In the polyatomic suite target, continuum and
MRMPS pseudo-continuum
orbitals are all represented by a linear combination of GTOs.
The target wave functions are expanded as a linear
combination of the configurations \emph{$\phi_k$}:
\begin{equation}
\Phi_i (\textbf{x}_1...\textbf{x}_N) = \sum_k c_{ik}\phi_k(\textbf{x}_1...\textbf{x}_N)\:,~\label{eq:targmod}
\end{equation}
where the \emph{c$_{ik}$} coefficients are determined by diagonalising the 
Hamiltonian of the molecular target.  
The quality of the target wave functions is dependent on the size of this 
expansion, as is indirectly, the quality of the scattering calculation 
\cite{jt189}.

The central idea of the MRMPS method is the augmentation of the
close-coupling expansion \eref{eq:rmat} with extra ``target'' wave
functions $\Phi_i$ that represent pseudo-states. Unlike the usual
target wave functions, these pseudo-states are not approximations to
true eigenstates of the target, but are used to represent a
discretised version of the electronic continuum.  These pseudo-states
are obtained by diagonalising the target electronic Hamiltonian
expressed in an appropriate basis of configurations, see below.

As we are considering an anionic target, there are a finite number
of bound target electronic states: three in the case C$_2^-$. This
means that, unlike previous MRMPS studies \cite{jt341,jt354}, the pseudo-states
are expected to all lie above the threshold to ionisation (electron
detachment) where they represent the discretised target continuum.
Of course, as the pseudo-continuum orbitals are added to the target
basis they also influence the representation of the target states.
It is a standard and tested assumption of the R-matrix with
pseudo-states method that electron
impact ionisation cross sections can be obtained by summing the cross
sections for excitation of pseudo-states which lie above the
vertical excitation threshold
\cite{bb96}. 

To generate configurations that describe an ionised target, the MRMPS
method uses and extra set of orbitals called pseudo-continuum orbitals
(PCOs). These orbitals are used to describe the ionised electron. The
PCOs are expanded in terms an even-tempered basis set \cite{EVEN} of
GTOs centred at the centre of mass of the system. In this type of
basis sets, the exponents of the GTOs follow:
\begin{equation}
\alpha_i=\alpha_0\beta^{(i-1)}, \qquad  \beta > 1, \qquad i=1,....,L\:,
\label{alphas}
\end{equation}
where by choosing different values of the parameters $\alpha_0$ and
$\beta$ different basis sets can be systematically generated. This is
useful for checking the convergence and stability of the calculation and,
of particular importance here, identifying physical resonances as
distinct from the pseudo-resonances which are a known artifact of the
RMPS procedure \cite{bhs96}. 

Given that the COs are also represented by GTOs, to avoid problems
with linear dependence it is necessary to remove those CO basis
functions from the set whose exponents are greater than $\alpha_0$. Even after
this condition care has to be taken to ensure that the three basis
sets (target, CO and PCO) give a linearly independent set of
orbitals. To do this the PCOs are first Schmidt orthogonalised to the
target molecular orbitals (MOs) and then symmetric orthogonalised
among themselves. At this stage those orbitals with eigenvalues of the
overlap matrix less than a deletion threshold $\delta$, here taken
as $4 \times
10^{-6}$, were assumed to be linearly dependent and were removed from the
basis.  This procedure is repeated for the COs, here using $\delta = 2
\times 10^{-6}$.

\section{Calculations}

\subsection{Target representation}

The starting point for the present calculations was our previous study
of the electron -- C$_2$ system \cite{jt383}. In that work the states
of C$_2$ were represented using the double-zeta plus polarisation
(DZP) Gaussian basis set of \citeasnoun{Dun70}, natural orbitals and
a complete active space configuration interaction (CAS-CI) 
in which the four 1s electrons were frozen in the
1\emph{$\sigma_g$} and 1\emph{$\sigma_u$} orbitals, and the
remaining eight electrons were freely distributed among the
2\emph{$\sigma_g$}, 3\emph{$\sigma_g$}, 2\emph{$\sigma_u$},
3\emph{$\sigma_u$}, 1\emph{$\pi_u$} and 1\emph{$\pi_g$} orbitals
giving configurations which can be written (1\emph{$\sigma_g$}
1\emph{$\sigma_u$})$^4$ (2\emph{$\sigma_g$} 3\emph{$\sigma_g$}
2\emph{$\sigma_u$} 3\emph{$\sigma_u$} 1\emph{$\pi_u$}
1\emph{$\pi_g$})$^8$. To treat C$_2^-$ we added
a PCO basis comprising 10 s, 10 p and 6 d orbitals. In this subsection
all results presented used PCO exponents generated using 
$\alpha_0$=0.17 and $\beta$=1.4.
The calculations presented here are for C$_2^-$
in its equilibrium geometry for which $R=2.396$~$a_0$ \cite{diatombook}.

The previous MRMPS studies considered electron impact ionisation of
H$_2$ and H$_3^+$ \cite{jt341,jt354}, which are both two electron
systems.  In this case the construction of ionised target plus PCO
configurations is straightforward. This is not so here where there are
many possible configurations that could be selected. The need to (a)
get a good representation of the (pseudo)-continuum, (b) get good
energies and wave functions for the physical states of the target, (c)
obtain a balanced description between the 
target and scattering calculations and (d)
keep the whole calculation computationally tractable meant that
considerable experimentation was required. \Tref{tab:targmod}
summarises the models, giving the target configurations used in
\eref{eq:targmod}, which were tested as part of the present study.

Besides choosing a  basis set and a  set of configurations, it is also
necessary to build a set  of target molecular orbitals. For  standard
scattering calculations  these orbitals are normally  ones
associated with the  $N$-electron target  under consideration. However
for calculations involving ionisation the final  target state only has
$N-1$ electrons; experience \cite{jt341}, echoed by tests performed as
part of this work, has shown that use  of orbitals associated with the
ionised target  gives best  results. We  therefore tested  two sets of
C$_2$ MOs, those generated from  a self consistent field  (SCF) calculation
and natural orbitals (NOs)   generated  using the  prescription  given
previously \cite{jt383}.

\begin{table}
\begin{center}
\caption{\label{tab:targmod} Configurations used in the various
target models tested. $N$ is the size of the resulting Hamiltonian
matrix for $^2$A$_g$
symmetry and PCO means pseudo continuum orbital.}

\begin{tabular}{crl}
\br
Model& N& Configurations\\
\mr
1 & 1110 &
  (1$\sigma_g$ 1$\sigma_u$)$^4$ (2$\sigma_g$ 3$\sigma_g$ 4$\sigma_g$
  2$\sigma_u$ 3$\sigma_u$ 1$\pi_u$ 1$\pi_g$)$^9$\\
&&(1$\sigma_g$ 2$\sigma_g$ 1$\sigma_u$ 2$\sigma_u$ 1$\pi_u$)$^{12}$ (PCOs)$^1$\\
2 & 140 &
  (1$\sigma_g$ 2$\sigma_g$ 1$\sigma_u$ 2$\sigma_u$)$^8$ 1$\pi_u$$^4$
  (3$\sigma_g$ 3$\sigma_u$ 1$\pi_g$)$^1$\\
&&(1$\sigma_g$ 2$\sigma_g$ 1$\sigma_u$ 2$\sigma_u$)$^8$ 1$\pi_u$$^3$
(3$\sigma_g$ 3$\sigma_u$ 1$\pi_g$)$^2$\\
&&(1$\sigma_g$ 2$\sigma_g$ 1$\sigma_u$ 2$\sigma_u$)$^8$ 1$\pi_u$$^4$ (PCOs)$^1$\\
&&(1$\sigma_g$ 2$\sigma_g$ 1$\sigma_u$ 2$\sigma_u$)$^8$ 1$\pi_u$$^3$
(3$\sigma_g$ 3$\sigma_u$ 1$\pi_g$)$^1$ (PCOs)$^1$\\
3 & 1600 &
(1$\sigma_g$ 1$\sigma_u$ 2$\sigma_g$)$^6$ 1$\pi_u$$^4$ (2$\sigma_u$ 
3$\sigma_g$ 3$\sigma_u$ 1$\pi_g$)$^3$\\
&&(1$\sigma_g$ 1$\sigma_u$ 2$\sigma_g$)$^6$ 1$\pi_u$$^3$ (2$\sigma_u$ 
3$\sigma_g$ 3$\sigma_u$ 1$\pi_g$)$^4$\\
&&(1$\sigma_g$ 1$\sigma_u$ 2$\sigma_g$)$^6$ 1$\pi_u$$^4$ (2$\sigma_u$ 
3$\sigma_g$ 3$\sigma_u$ 1$\pi_g$)$^2$ (PCOs)$^1$\\
&&(1$\sigma_g$ 1$\sigma_u$ 2$\sigma_g$)$^6$ 1$\pi_u$$^3$ (2$\sigma_u$ 
3$\sigma_g$ 3$\sigma_u$ 1$\pi_g$)$^3$ (PCOs)$^1$\\
4 & 425 &
(1$\sigma_g$ 1$\sigma_u$)$^4$ (2$\sigma_g$ 3$\sigma_g$ 2$\sigma_u$
3$\sigma_u$ 1$\pi_u$ 1$\pi_g$)$^9$\\
&&(1$\sigma_g$ 2$\sigma_g$ 1$\sigma_u$ 2$\sigma_u$)$^8$ 1$\pi_u$$^4$ (PCOs)$^1$\\
&&(1$\sigma_g$ 2$\sigma_g$ 1$\sigma_u$ 2$\sigma_u$)$^8$ 1$\pi_u$$^3$
(3$\sigma_g$ 3$\sigma_u$ 1$\pi_g$)$^1$ (PCOs)$^1$\\
5 & 597 &
(1$\sigma_g$ 1$\sigma_u$)$^4$ (2$\sigma_g$ 3$\sigma_g$ 2$\sigma_u$
3$\sigma_u$ 1$\pi_u$ 1$\pi_g$)$^9$\\
&&(1$\sigma_g$ 1$\sigma_u$)$^4$ (2$\sigma_g$ 2$\sigma_u$ 1$\pi_u$)$^8$ 
(PCOs)$^1$\\
&&(1$\sigma_g$ 1$\sigma_u$)$^4$ (2$\sigma_g$ 2$\sigma_u$ 1$\pi_u$)$^7$
(3$\sigma_g$ 3$\sigma_u$ 1$\pi_g$)$^1$ (PCOs)$^1$\\
6 & 3111 &
(1$\sigma_g$ 1$\sigma_u$)$^4$ (2$\sigma_g$ 3$\sigma_g$ 2$\sigma_u$
3$\sigma_u$ 1$\pi_u$ 1$\pi_g$)$^9$\\
&&(1$\sigma_g$ 1$\sigma_u$)$^4$ (2$\sigma_g$ 2$\sigma_u$ 1$\pi_u$)$^8$ 
(PCOs)$^1$\\
&&(1$\sigma_g$ 1$\sigma_u$)$^4$ (2$\sigma_g$ 2$\sigma_u$ 1$\pi_u$)$^7$
(3$\sigma_g$ 3$\sigma_u$ 1$\pi_g$)$^1$ (PCOs)$^1$\\
&&(1$\sigma_g$ 1$\sigma_u$)$^4$ (2$\sigma_g$ 2$\sigma_u$ 1$\pi_u$)$^6$
(3$\sigma_g$ 3$\sigma_u$ 1$\pi_g$)$^2$ (PCOs)$^1$\\
7 & 20454 & 
(1$\sigma_g$ 1$\sigma_u$ 2$\sigma_g$)$^6$ (2$\sigma_u$ 3$\sigma_g$ 
3$\sigma_u$ 1$\pi_g$ 1$\pi_u$)$^7$\\
&&(1$\sigma_g$ 1$\sigma_u$ 2$\sigma_g$)$^6$ (2$\sigma_u$ 3$\sigma_g$ 
3$\sigma_u$ 1$\pi_g$ 1$\pi_u$)$^6$ (PCOs)$^1$\\
8 & 97500 &
(1$\sigma_g$ 1$\sigma_u$)$^4$ (2$\sigma_g$ 3$\sigma_g$ 4$\sigma_g$
2$\sigma_u$ 3$\sigma_u$ 1$\pi_u$ 1$\pi_g$)$^9$\\
&&(1$\sigma_g$ 1$\sigma_u$)$^4$ (2$\sigma_g$ 3$\sigma_g$ 4$\sigma_g$
2$\sigma_u$ 3$\sigma_u$ 1$\pi_u$ 1$\pi_g$)$^8$ (PCOs)$^1$\\
\br
\end{tabular}
\end{center}
\end{table}

\Tref{tab:tarE} gives energies for states of C$_2^-$ calculated
with the various models and orbitals sets. These can be compared with the
results of our previous study \cite{jt383} which, once nuclear motion
effects had been taken into account, gave good agreement with the experimental
results reported in \citeasnoun{diatombook}. 

Considering each model in turn. Model 1 gave reasonable target
energies but the pseudo-state configurations are very limited as they
are obtain from a single electron in a PCO and a frozen target C$_2$;
this treatment is not consistent with the CAS-CI representation of
C$_2^-$. Model 2 is more consistent in that 8 electrons are frozen in
all configurations. However this much more limited model gave poor
energies for the target states; indeed with SCF MOs it did not even
predict the correct ground state for C$_2^-$. Model 3 is built on
Model 2 but with only 6 target electrons completely frozen; it gives
rather a large scattering Hamiltonian and when tested did not
give particularly good eigenphase sums. Model 4 was built as a hybrid
between Models 1 and 2 using the target configurations from the former
and pseudo-state configurations from the latter. This model gave good
target energies and a relatively small Hamiltonian size for the
scattering problem. This model formed the basis of our
preliminary study \cite{jt450} and became our workhorse for test calculations.
Model 5 is a slightly enlarged version of model 4 and behaves similarly.

Use of the partitioned R-matrix method allowed us to explore target
models which implied significantly larger Hamiltonians for the
scattering problem.  We therefore tested the effect of gradually
expanding the CAS-CI used to generate the pseudo-states. The limit of
the process is Model 8, in which the same extended CAS is used in both
parts of the calculation. However, its corresponding scattering model
is far too big to be tractable (see \tref{tab:scatmod}) and
was not pursued. Our attempts to construct intermediate models,
gives numbers 6 and 7. Model 7 did not give good target energies. We therefore
decided to concentrate on use of models 4,  5 and 6.

As can be seen from \tref{tab:tarE}, the use of C$_2$ SCF MOs gives
results of similar quality to NOs, in contrast to more usual
calculations were NOs give significantly better
results. This is perhaps not surprising since
the NOs were constructed to give a good representation of a range of
electronically excited states of C$_2$ which is not our purpose here.
The calculations reported below use SCF MOs unless otherwise stated.

\begin{table}
\begin{center}
\caption{\label{tab:tarE} C$_2$$^-$ ground state energies 
(negative numbers in E$_h$)
and excitation energies (in eV) for different models and orbitals.}
\begin{tabular}{ccccccccc}
\br
Model  &\multicolumn{2}{c}{X $^2\Sigma_\emph{g}^+$}&
&\multicolumn{2}{c}{A $^2\Pi_\emph{u}$}&
&\multicolumn{2}{c}{B $^2\Sigma_\emph{u}^+$}\\
\cline{2-3} \cline{5-6} \cline{8-9}
 &SCF MOs&C$_2$ NOs& &SCF MOs&C$_2$ NOs& &SCF MOs&C$_2$ NOs\\
\mr
1&-75.61166&-& &0.729&-& &2.673&-\\
2&1.951&-75.53709& &-75.56659&0.850& &7.955&5.813\\
3&-75.59480&-75.58782& &0.597&0.611& &3.115&2.549\\
4&-75.60956&-75.60944& &0.468&0.419& &2.734&2.506\\
5&-75.62675&-& &0.883&-& &2.717&-\\
6&-75.65777&-& &0.912&-& &2.908&-\\
7&0.123&-75.71419& &-75.66055&0.686& &1.823&2.621\\
8&-75.72075& & &0.689& & &2.621& \\
$a$& & -75.67213 & & & 0.557 & & & 2.355\\
\br
\end{tabular}
\end{center}
$^a$ Previous calculations  \cite{jt383}.
\end{table}

There is one further target property which proved to be of considerable
importance for the scattering calculations, that is the long range
polarisability of C$_2^-$ target. \citeasnoun{jt341} showed that the MRMPS
method converges the long-range target polarisability in a way that
standard close-coupling expansions do not. As is discussed 
below, a good representation of the polarisability is essential to
get a good physical model of the electron -- molecular anion collision
as it provides the dominant attractive term in the interaction.

\Tref{tab:polar} shows the isotropic polarisabilities predicted by
various of our target models. We could find no literature value for this
parameter. Indeed it is not straightforward to calculate it with standard
electronic structure codes, we tried, as they are not generally set up
to treat molecular anions. The table shows that the 
calculated polarisability is
generally stable with changes to the PCO basis but not with other aspects of
the target model. In part this is because both excited states of C$_2^-$
are dipole allowed from the ground state and low-lying. This makes the
calculated polarisability particularly sensitive to the transition
dipole and precise excitation energies of these states. Our results suggest
that the true polarisabilty of C$_2^-$ at $R=2.396$~$a_0$ is between 25
and 32 a$_0^3$, of this less than 10 a$_0^3$ arises from coupling to
the two physical electronically excited states of C$_2^-$.

\begin{table}
\caption{\label{tab:polar}  Isotropic
polarisabilities of C$_2^-$ in a$_0^3$ 
for different models, orbitals and PCO basis determined by
($\alpha_0$,$\beta$).}
\begin{center}
\begin{tabular}{ccccccc}
\br
Model &\multicolumn{2}{c}{$\alpha_0$=0.17 $\beta$=1.4}&$\alpha_0$=0.15 $\beta$=1.4&$\alpha_0$=0.17 $\beta$=1.5&\multicolumn{2}{c}{$\alpha_0$=0.17 $\beta$=1.3}\\ \cline{2-3} \cline{2-3} \cline{6-7}
 &SCF MOs&C$_2$ NOs&SCF MOs&SCF MOs&SCF MOs&C$_2$ NOs\\
\mr
 1 &26.42& & & & & \\
 2&28.58&109.51& & & & \\
 3&67.49&12.24& & & & \\
 4&32.24& &32.48&31.24&31.98&18.80\\
 5 &25.00& & & & & \\
 6 &24.55& & & & & \\
\br
\end{tabular}
\end{center}
\end{table}

\begin{figure}
\begin{center}
{\resizebox{130mm}{!}{ \includegraphics*{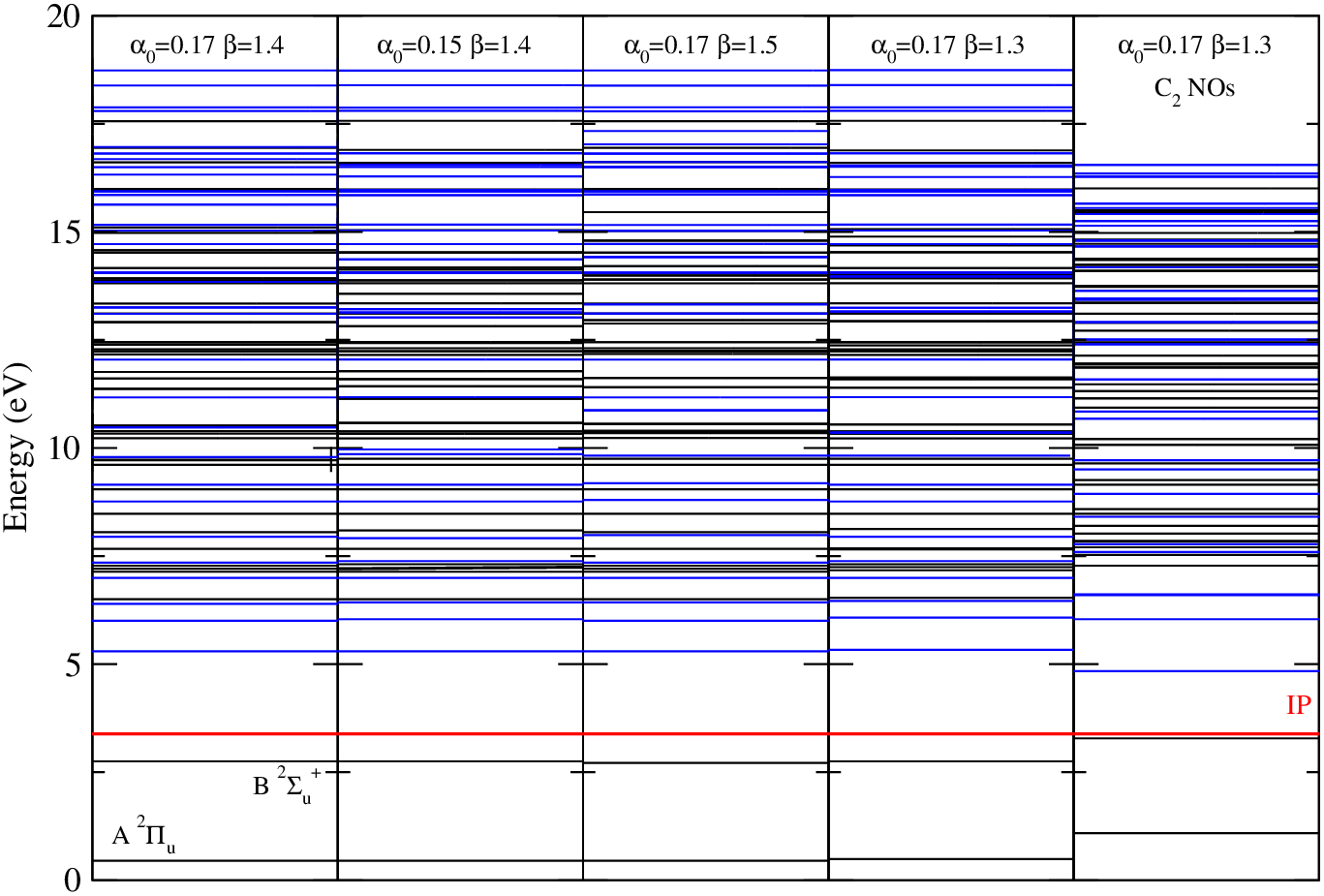}}}
\caption{Target state distribution for C$_2^-$ for Model~4 for various 
($\alpha_0$,$\beta$) values as indicated in the figure. 
Energies are relative to the X~$^3\Sigma_g^+$ ground state of C$_2^-$.
SCF MOs
of C$_2$ were used  except where otherwise indicated. The horizontal
line marked IP denotes the ionisation threshold.}
\label{fig:enlev}
\end{center}
\end{figure}

So far we have concentrated on our calculations of the physical
properties of the target. There is however one other property of
importance for the scattering runs: the spectral coverage of the
continuum by the pseudo-states. \Fref{fig:enlev} compares energy
levels generated by various versions of Model 4. The behaviour shown
here, that the energy levels are stable to choice of PCO basis but
very sensitive to the target model used, was also shown in the other
comparisons we made
\cite{gthesis}. A common feature of all the comparisons is the
sparsity of states directly above ionisation. In principle one could
get pseudo-states in this energy region by 
significantly expanding the R-matrix box size
and the associated PCO basis; such a calculation was not
deemed computationally tractable at present 
However, as can be seen from the
results presented below, our pseudo-state distributions
lead to very small near-threshold electron impact
detachment cross sections, which appears to be in
agreement with the observations.

\subsection{Scattering model}

Test calculations using Model~1 for R-matrix spheres of radius
$a=10$~a$_0$ and 13~a$_0$ showed that $a=10$~a$_0$ gave stable
results and this was used for all further
calculations. CO's were taken from
\citeasnoun{jt286} with the largest exponent functions removed. This basis
contains functions with $\ell$ up to 4 ie g orbitals. Use of two 
different sets of PCOs were tested:\\
4-14$a_g$ 2-7$b_{2u}$ 2-7$b_{3u}$ 1-4$b_{1g}$ 4-8$b_{1u}$
2-5$b_{3g}$ 2-5$b_{2g}$ 1$a_u$\\
4-24$a_g$ 2-10$b_{2u}$ 2-10$b_{3u}$ 1-7$b_{1g}$ 4-12$b_{1u}$
2-8$b_{3g}$ 2-8$b_{2g}$ 1$a_u$\\
in D$_{2h}$ notation. The eigenphase calculated with the larger PCO set 
was slightly higher, so this was used for further calculation. 

To go from the target models detailed above to the inner region R-matrix
wave functions of \eref{eq:rmat} we followed the prescription
developed previously \cite{jt189} for standard R-matrix calculations.
This prescription, which has been used successfully in many studies,
is designed to provide a balanced treatment between the target, $N$-electron,
and scattering, $N+1$-electron problems. \Tref{tab:scatmod} details
the configurations generated for the $N+1$-electron calculations
associated with each of the models described above. Also given is the
size of the final Hamiltonian, a crucial parameter in determining the
tractability of the calculation. Although our Hamiltonian construction
algorithm is very efficient meaning the computational demands of this
step of the calculation only depend weakly on the number of target states
(and pseudo-states) included in the expansion \cite{jt180}, the
diagonalisation step still represents a major bottleneck. As a result, the
largest previously published molecular R-matrix calculation was restricted
to $N \sim 28000$ \cite{jt323}.

\begin{table}
{\scriptsize{
\begin{center}
\caption{\label{tab:scatmod} Configurations used for the various
scattering models tested. $N$ is the size of the resulting Hamiltonian
matrix with $^1$A$_g$ symmetry
and PCO means pseudo continuum orbital and CO means continuum
orbital.}

\begin{tabular}{crl}
\br
Model& N& Configurations\\
\mr
1 & 21705 &
(1$\sigma_g$ 1$\sigma_u$)$^4$ (2$\sigma_g$ 3$\sigma_g$ 4$\sigma_g$
2$\sigma_u$ 3$\sigma_u$ 1$\pi_u$ 1$\pi_g$)$^9$ (COs)$^1$\\
&&(1$\sigma_g$ 2$\sigma_g$ 1$\sigma_u$ 2$\sigma_u$ 1$\pi_u$)$^{12}$ 
(PCOs)$^1$ (COs)$^1$\\
&&(1$\sigma_g$ 1$\sigma_u$)$^4$ (2$\sigma_g$ 3$\sigma_g$ 4$\sigma_g$
2$\sigma_u$ 3$\sigma_u$ 1$\pi_u$ 1$\pi_g$)$^{10}$\\
&&(1$\sigma_g$ 2$\sigma_g$ 1$\sigma_u$ 2$\sigma_u$ 1$\pi_u$)$^{12}$
(PCOs)$^2$\\
&&(1$\sigma_g$ 2$\sigma_g$ 1$\sigma_u$ 2$\sigma_u$ 1$\pi_u$)$^{12}$
(3$\sigma_g$ 4$\sigma_g$ 3$\sigma_u$ 1$\pi_g$)$^1$ (PCOs)$^1$\\
2 & 6447 &
(1$\sigma_g$ 2$\sigma_g$ 1$\sigma_u$ 2$\sigma_u$)$^8$ 1$\pi_u$$^4$
(3$\sigma_g$ 3$\sigma_u$ 1$\pi_g$)$^1$ (COs)$^1$\\
&&(1$\sigma_g$ 2$\sigma_g$ 1$\sigma_u$ 2$\sigma_u$)$^8$ 1$\pi_u$$^3$
(3$\sigma_g$ 3$\sigma_u$ 1$\pi_g$)$^2$ (COs)$^1$\\
&&(1$\sigma_g$ 2$\sigma_g$ 1$\sigma_u$ 2$\sigma_u$)$^8$ 1$\pi_u$$^4$ 
(PCOs)$^1$ (COs)$^1$\\
&&(1$\sigma_g$ 2$\sigma_g$ 1$\sigma_u$ 2$\sigma_u$)$^8$ 1$\pi_u$$^3$
(3$\sigma_g$ 3$\sigma_u$ 1$\pi_g$)$^1$ (PCOs)$^1$ (COs)$^1$\\
&&(1$\sigma_g$ 2$\sigma_g$ 1$\sigma_u$ 2$\sigma_u$)$^8$ 1$\pi_u$$^4$
(3$\sigma_g$ 3$\sigma_u$ 1$\pi_g$)$^2$\\
&&(1$\sigma_g$ 2$\sigma_g$ 1$\sigma_u$ 2$\sigma_u$)$^8$ 1$\pi_u$$^4$ 
(3$\sigma_g$ 3$\sigma_u$ 1$\pi_g$)$^1$ (PCOs)$^1$\\
&&(1$\sigma_g$ 2$\sigma_g$ 1$\sigma_u$ 2$\sigma_u$$^8$ 1$\pi_u$$^4$ 
(PCOs)$^2$\\
&&(1$\sigma_g$ 2$\sigma_g$ 1$\sigma_u$ 2$\sigma_u$)$^8$ 1$\pi_u$$^3$ 
(3$\sigma_g$ 3$\sigma_u$ 1$\pi_g$)$^2$ (PCOs)$^1$\\
&&(1$\sigma_g$ 2$\sigma_g$ 1$\sigma_u$ 2$\sigma_u$)$^8$ 1$\pi_u$$^3$ 
(3$\sigma_g$ 3$\sigma_u$ 1$\pi_g$)$^1$ (PCOs)$^2$\\
3 & 52270 &
(1$\sigma_g$ 1$\sigma_u$ 2$\sigma_g$)$^6$ 1$\pi_u$$^4$ (2$\sigma_u$
3$\sigma_g$ 3$\sigma_u$ 1$\pi_g$)$^3$ (COs)$^1$\\
&&(1$\sigma_g$ 1$\sigma_u$ 2$\sigma_g$)$^6$ 1$\pi_u$$^3$ (2$\sigma_u$
3$\sigma_g$ 3$\sigma_u$ 1$\pi_g$)$^4$ (COs)$^1$\\
&&(1$\sigma_g$ 1$\sigma_u$ 2$\sigma_g$)$^6$ 1$\pi_u$$^4$ (2$\sigma_u$
3$\sigma_g$ 3$\sigma_u$ 1$\pi_g$)$^2$ (PCOs)$^1$ (COs)$^1$\\
&&(1$\sigma_g$ 1$\sigma_u$ 2$\sigma_g$)$^6$ 1$\pi_u$$^3$ (2$\sigma_u$
3$\sigma_g$ 3$\sigma_u$ 1$\pi_g$)$^3$ (PCOs)$^1$ (COs)$^1$\\
&&(1$\sigma_g$ 1$\sigma_u$ 2$\sigma_g$)$^6$ 1$\pi_u$$^4$ (2$\sigma_u$
3$\sigma_g$ 3$\sigma_u$ 1$\pi_g$)$^4$\\
&&(1$\sigma_g$ 1$\sigma_u$ 2$\sigma_g$)$^6$ 1$\pi_u$$^4$ (2$\sigma_u$
3$\sigma_g$ 3$\sigma_u$ 1$\pi_g$)$^3$ (PCOs)$^1$\\
&&(1$\sigma_g$ 1$\sigma_u$ 2$\sigma_g$)$^6$ 1$\pi_u$$^4$ (2$\sigma_u$
3$\sigma_g$ 3$\sigma_u$ 1$\pi_g$)$^2$ (PCOs)$^2$\\
&&(1$\sigma_g$ 1$\sigma_u$ 2$\sigma_g$)$^6$ 1$\pi_u$$^3$ (2$\sigma_u$
3$\sigma_g$ 3$\sigma_u$ 1$\pi_g$)$^4$ (PCOs)$^1$\\
&&(1$\sigma_g$ 1$\sigma_u$ 2$\sigma_g$)$^6$ 1$\pi_u$$^3$ (2$\sigma_u$
3$\sigma_g$ 3$\sigma_u$ 1$\pi_g$)$^3$ (PCOs)$^2$\\
4 & 6575 &
(1$\sigma_g$ 1$\sigma_u$)$^4$ (2$\sigma_g$ 3$\sigma_g$ 2$\sigma_u$
3$\sigma_u$ 1$\pi_u$ 1$\pi_g$)$^9$ (COs)$^1$\\
&&(1$\sigma_g$ 2$\sigma_g$ 1$\sigma_u$ 2$\sigma_u$)$^8$ 1$\pi_u$$^4$ 
(PCOs)$^1$ (COs)$^1$\\
&&(1$\sigma_g$ 2$\sigma_g$ 1$\sigma_u$ 2$\sigma_u$)$^8$ 1$\pi_u$$^3$
(3$\sigma_g$ 3$\sigma_u$ 1$\pi_g$)$^1$ (PCOs)$^1$ (COs)$^1$\\
&&(1$\sigma_g$ 1$\sigma_u$)$^4$ (2$\sigma_g$ 3$\sigma_g$ 2$\sigma_u$
3$\sigma_u$ 1$\pi_u$ 1$\pi_g$)$^{10}$\\
&&(1$\sigma_g$ 2$\sigma_g$ 1$\sigma_u$ 2$\sigma_u$)$^8$ 1$\pi_u$$^4$ 
(PCOs)$^2$\\
&&(1$\sigma_g$ 2$\sigma_g$ 1$\sigma_u$ 2$\sigma_u$)$^8$ 1$\pi_u$$^3$ 
(3$\sigma_g$ 3$\sigma_u$ 1$\pi_g$)$^2$ (PCOs)$^1$\\
&&(1$\sigma_g$ 2$\sigma_g$ 1$\sigma_u$ 2$\sigma_u$)$^8$ 1$\pi_u$$^3$ 
(3$\sigma_g$ 3$\sigma_u$ 1$\pi_g$)$^1$ (PCOs)$^2$\\
5 & 12283 &
(1$\sigma_g$ 1$\sigma_u$)$^4$ (2$\sigma_g$ 3$\sigma_g$ 2$\sigma_u$
3$\sigma_u$ 1$\pi_u$ 1$\pi_g$)$^9$ (COs)$^1$\\
&&(1$\sigma_g$ 1$\sigma_u$)$^4$ (2$\sigma_g$ 2$\sigma_u$ 1$\pi_u$)$^8$ 
(PCOs)$^1$ (COs)$^1$\\
&&(1$\sigma_g$ 1$\sigma_u$)$^4$ (2$\sigma_g$ 2$\sigma_u$ 1$\pi_u$)$^7$ 
(3$\sigma_g$ 3$\sigma_u$ 1$\pi_g$)$^1$ (PCOs)$^1$ (COs)$^1$\\
&&(1$\sigma_g$ 1$\sigma_u$)$^4$ (2$\sigma_g$ 3$\sigma_g$ 2$\sigma_u$
3$\sigma_u$ 1$\pi_u$ 1$\pi_g$)$^{10}$\\
&&(1$\sigma_g$ 1$\sigma_u$)$^4$ (2$\sigma_g$ 2$\sigma_u$ 1$\pi_u$)$^8$
(PCOs)$^2$\\
&&(1$\sigma_g$ 1$\sigma_u$)$^4$ (2$\sigma_g$ 2$\sigma_u$ 1$\pi_u$)$^7$
(3$\sigma_g$ 3$\sigma_u$ 1$\pi_g$)$^2$ (PCOs)$^1$\\
&&(1$\sigma_g$ 1$\sigma_u$)$^4$ (2$\sigma_g$ 2$\sigma_u$ 1$\pi_u$)$^7$
(3$\sigma_g$ 3$\sigma_u$ 1$\pi_g$)$^1$ (PCOs)$^2$\\
6 & 12283 &
(1$\sigma_g$ 1$\sigma_u$)$^4$ (2$\sigma_g$ 3$\sigma_g$ 2$\sigma_u$
3$\sigma_u$ 1$\pi_u$ 1$\pi_g$)$^9$ (COs)$^1$\\
&&(1$\sigma_g$ 1$\sigma_u$)$^4$ (2$\sigma_g$ 2$\sigma_u$ 1$\pi_u$)$^8$ 
(PCOs)$^1$ (COs)$^1$\\
&&(1$\sigma_g$ 1$\sigma_u$)$^4$ (2$\sigma_g$ 2$\sigma_u$ 1$\pi_u$)$^7$ 
(3$\sigma_g$ 3$\sigma_u$ 1$\pi_g$)$^1$ (PCOs)$^1$ (COs)$^1$\\
&&(1$\sigma_g$ 1$\sigma_u$)$^4$ (2$\sigma_g$ 2$\sigma_u$ 1$\pi_u$)$^6$ 
(3$\sigma_g$ 3$\sigma_u$ 1$\pi_g$)$^2$ (PCOs)$^1$ (COs)$^1$\\
&&(1$\sigma_g$ 1$\sigma_u$)$^4$ (2$\sigma_g$ 3$\sigma_g$ 2$\sigma_u$
3$\sigma_u$ 1$\pi_u$ 1$\pi_g$)$^{10}$\\
&&(1$\sigma_g$ 1$\sigma_u$)$^4$ (2$\sigma_g$ 2$\sigma_u$ 1$\pi_u$)$^8$ 
(PCOs)$^2$\\
&&(1$\sigma_g$ 1$\sigma_u$)$^4$ (2$\sigma_g$ 2$\sigma_u$ 1$\pi_u$)$^7$ 
(3$\sigma_g$ 3$\sigma_u$ 1$\pi_g$)$^2$ (PCOs)$^1$\\
&&(1$\sigma_g$ 1$\sigma_u$)$^4$ (2$\sigma_g$ 2$\sigma_u$ 1$\pi_u$)$^7$ 
(3$\sigma_g$ 3$\sigma_u$ 1$\pi_g$)$^1$ (PCOs)$^2$\\
7 & 823823 &
(1$\sigma_g$ 1$\sigma_u$ 2$\sigma_g$)$^6$ (2$\sigma_u$ 3$\sigma_g$ 
3$\sigma_u$ 1$\pi_g$ 1$\pi_u$)$^7$ (COs)$^1$\\
&&(1$\sigma_g$ 1$\sigma_u$ 2$\sigma_g$)$^6$ (2$\sigma_u$ 3$\sigma_g$ 
3$\sigma_u$ 1$\pi_g$ 1$\pi_u$)$^6$ (PCOs)$^1$ (COs)$^1$\\
&&(1$\sigma_g$ 1$\sigma_u$ 2$\sigma_g$)$^6$ (2$\sigma_u$ 3$\sigma_g$ 
3$\sigma_u$ 1$\pi_g$ 1$\pi_u$)$^8$\\
&&(1$\sigma_g$ 1$\sigma_u$ 2$\sigma_g$)$^6$ (2$\sigma_u$ 3$\sigma_g$ 
3$\sigma_u$ 1$\pi_g$ 1$\pi_u$)$^7$ (PCOs)$^1$\\
&&(1$\sigma_g$ 1$\sigma_u$ 2$\sigma_g$)$^6$ (2$\sigma_u$ 3$\sigma_g$ 
3$\sigma_u$ 1$\pi_g$ 1$\pi_u$)$^6$ (PCOs)$^2$\\
&&(1$\sigma_g$ 1$\sigma_u$)$^4$ (2$\sigma_g$ 3$\sigma_g$ 2$\sigma_u$
3$\sigma_u$ 1$\pi_u$ 1$\pi_g$)$^{10}$\\
&&(1$\sigma_g$ 1$\sigma_u$)$^4$ (2$\sigma_g$ 2$\sigma_u$ 1$\pi_u$)$^8$ 
(PCOs)$^2$\\
&&(1$\sigma_g$ 1$\sigma_u$)$^4$ (2$\sigma_g$ 2$\sigma_u$ 1$\pi_u$)$^7$ 
(3$\sigma_g$ 3$\sigma_u$ 1$\pi_g$)$^2$ (PCOs)$^1$\\
&&(1$\sigma_g$ 1$\sigma_u$)$^4$ (2$\sigma_g$ 2$\sigma_u$ 1$\pi_u$)$^7$ 
(3$\sigma_g$ 3$\sigma_u$ 1$\pi_g$)$^1$ (PCOs)$^2$\\
8 & too big &
(1$\sigma_g$ 1$\sigma_u$)$^4$ (2$\sigma_g$ 3$\sigma_g$ 4$\sigma_g$
2$\sigma_u$ 3$\sigma_u$ 1$\pi_u$ 1$\pi_g$)$^9$ (COs)$^1$\\
&&(1$\sigma_g$ 1$\sigma_u$)$^4$ (2$\sigma_g$ 3$\sigma_g$ 4$\sigma_g$
2$\sigma_u$ 3$\sigma_u$ 1$\pi_u$ 1$\pi_g$)$^8$ (PCOs)$^1$ (COs)$^1$\\
&&(1$\sigma_g$ 1$\sigma_u$)$^4$ (2$\sigma_g$ 3$\sigma_g$ 4$\sigma_g$
2$\sigma_u$ 3$\sigma_u$ 1$\pi_u$ 1$\pi_g$)$^{10}$\\
&&(1$\sigma_g$ 1$\sigma_u$)$^4$ (2$\sigma_g$ 3$\sigma_g$ 4$\sigma_g$
2$\sigma_u$ 3$\sigma_u$ 1$\pi_u$ 1$\pi_g$)$^9$ (PCOs)$^1$\\
&&(1$\sigma_g$ 1$\sigma_u$)$^4$ (2$\sigma_g$ 3$\sigma_g$ 4$\sigma_g$
2$\sigma_u$ 3$\sigma_u$ 1$\pi_u$ 1$\pi_g$)$^8$ (PCOs)$^2$\\
\br
\end{tabular}
\end{center}
}}
\end{table}

Model 8 gives an extremely large final step Hamiltonians and no
attempt was made to actually solve for it.  Many calculations were
performed for Models 1 to 6 for testing purposes.
Calculations  using the partitioned R-matrix method
\cite{jt332} explicitly considered the lowest 1000 solutions which
were found to be sufficient to span scattering energies up to about
40~eV.

Scattering calculations were performed for both singlet and triplet
symmetries.  After some experimentation it was decided to include 114
target states in the final close-coupling expansion. This number is
evaluated in D$_{2h}$ symmetry so counts states which are degenerate
in D$_{\infty h}$ symmetry twice.  66 of these states are doublets of
which the lowest 3 are the physical states of C$_2^-$. The 48 quartet
states only couple to triplet calculations. These states span energies
up to about 19~eV above the target ground state for both Models 4, 5  and 6.

Outer region calculations were performed by propagating the R-matrices
to 100~a$_0$ and then matching to asymptotic Coulomb functions.
This large number of states makes
the outer calculations slow compared to standard R-matrix calculations but
we were still able to generate results for sufficient energies to map out
the various resonance structures, both real and pseudo.

\subsection{Born Approximation}

The interaction between the scattering electron and the C$_2^-$ anion
is long-range and repulsive. It can therefore be expected that many of
the collisions will occur with a large impact parameter. Such
collisions are not well represented in the calculations described
above, all of which use a CO basis set and hence partial wave expansion
truncated at $\ell =4$.  However collisions with high values of
electron angular momentum ($\ell > 4$) are amenable to a much simpler
treatment since in these collisions the scattering electron 
can be assumed not to penetrate the
charge cloud of the C$_2^-$ target. Under these circumstances the
collision cross sections are determined by purely long-range effects.

To allow the important effect of collisions with $\ell > 4$ in the
inelastic cross sections presented here, we used a simple top-up
procedure based on the Born approximation \cite{jt267}. There are a
number of ways applying such a Born top-up procedure but since C$_2^-$
is non-polar, meaning rotational effects will be small, and the
non-resonant cross sections at low $\ell$ are very small, meaning that
interference effects for these $\ell$'s will not be important, the
simplest procedure based on the direct augmentation of the cross
sections was used.  This procedure was applied to electron impact
electronic excitation of all physical and pseudo states which are
dipole allowed, that is to all excitations to states of
$^2\Sigma_\emph{u}^+$ and $^2\Pi_\emph{u}$ symmetry. Only transition
dipole moments were considered for the long-range potential. For the
pseudo-states this extra excitation cross sections were taken as
supplementing the electron impact detachment cross sections. This
increase turns out to be substantial showing that the majority of this
process does in fact occur through long-range collisions.

\section{Results}

A very large number of different calculations were attempted. For space
reasons only only those performed using Models 4, 5 and 6 will be reported in
detail. As a major aim of this study was to characterise the resonances
observed in the storage ring experiments, it is this aspect of the work we
discuss first. We will then give our results for total cross sections for
both electron impact electronic excitation and electron detachment.

\subsection{Resonances}

\begin{figure}
\begin{center}
{\resizebox{130mm}{!}{ \includegraphics*{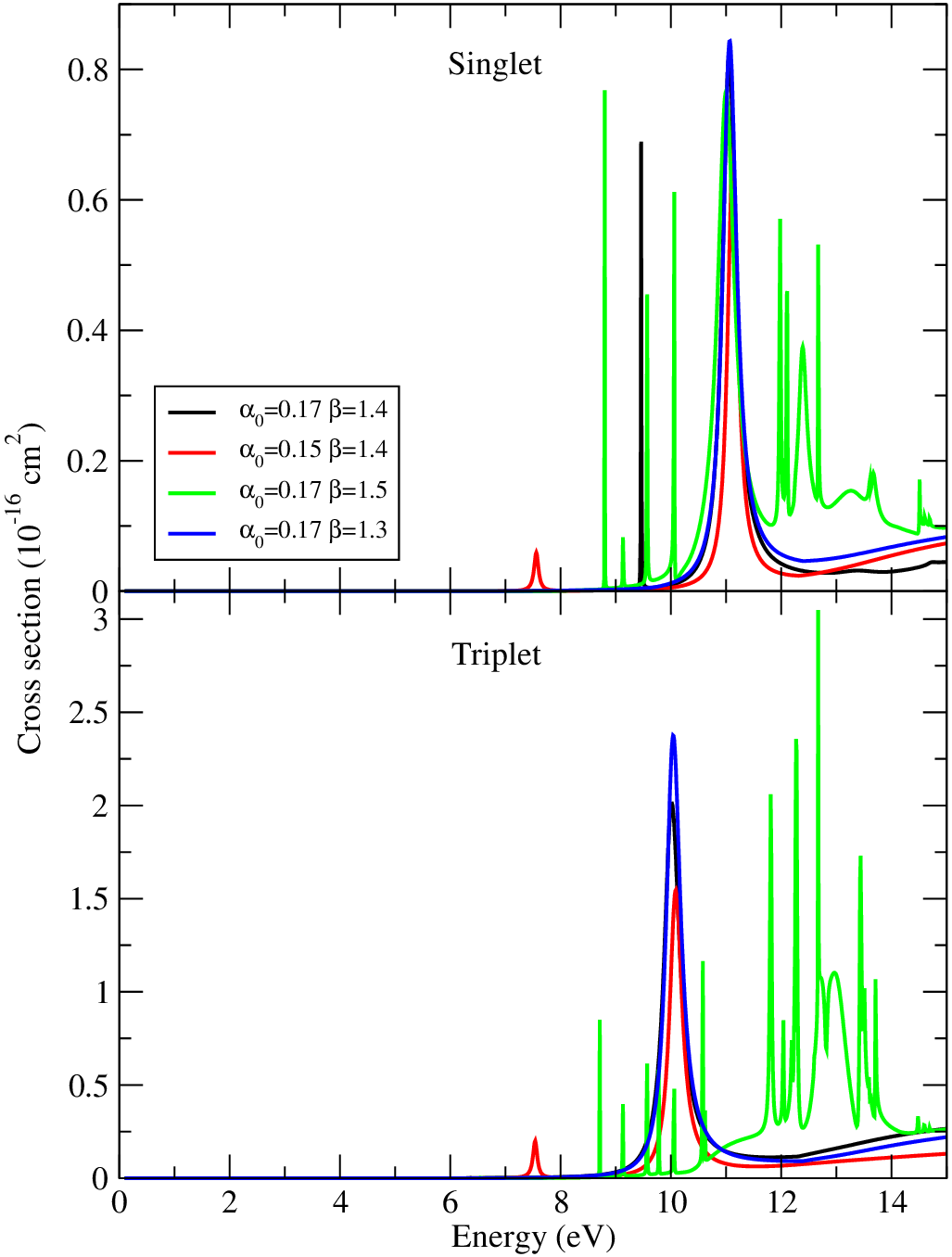}}}
\caption{Partial electron impact
detachment cross sections of C$_2^-$ for Model~4 for final
states of $^2$B$_{3g}$ ($^2\Pi_g$) total symmetry
with different values of ($\alpha_0$,$\beta$); ionisation to
singlet and triplet final states are shown to illustrate their
distinct resonance structures.}
\label{fig:2B3g}
\end{center}
\end{figure}

All models tested showed a $^1\Sigma_g^+$ resonance feature at about
5~eV. However only the more sophisticated models showed stable resonances
in the energy region probed by the storage ring experiments.

\begin{figure}
\begin{center}
{\resizebox{130mm}{!}{ \includegraphics*{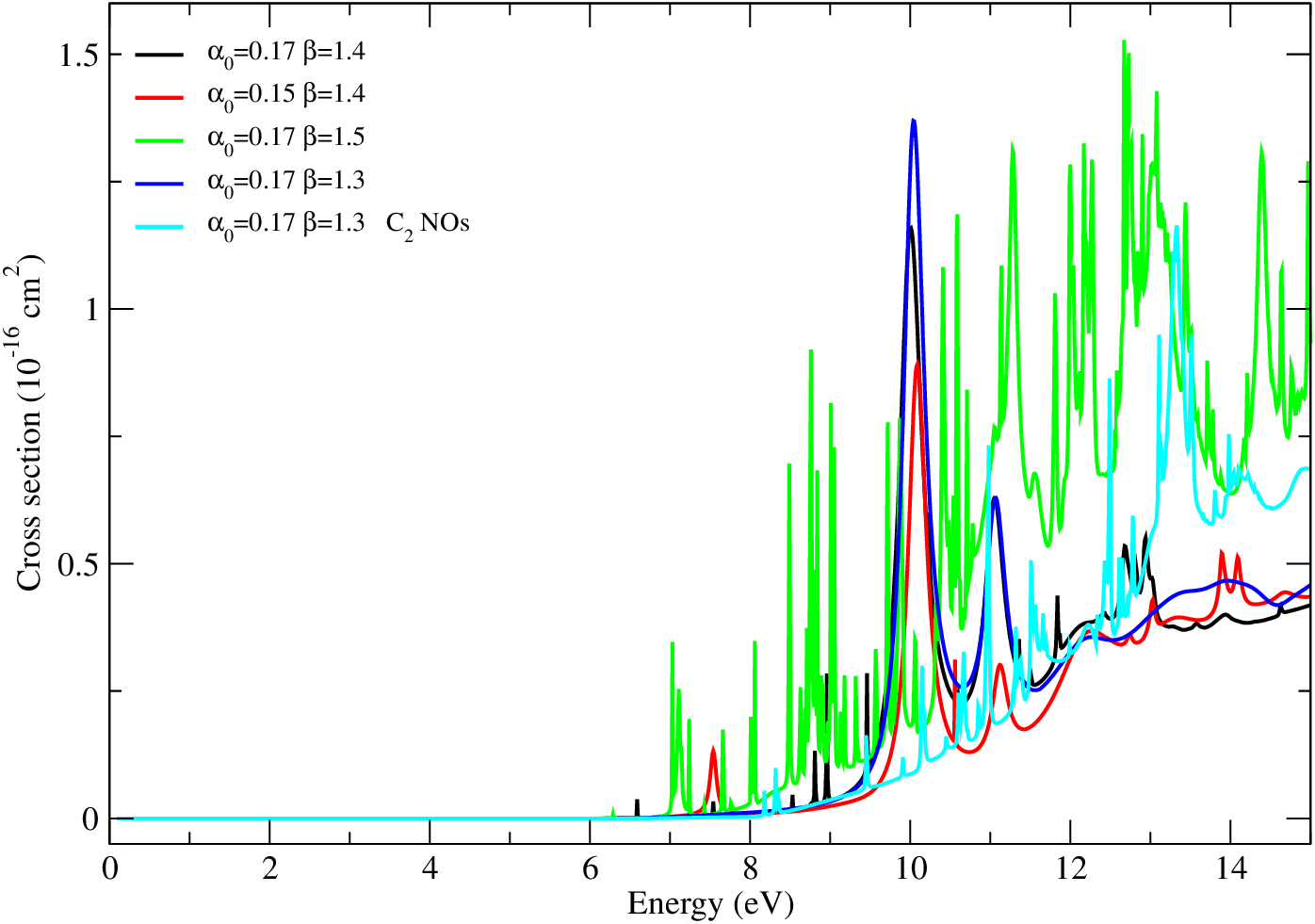}}}
\caption{Total electron impact detachment cross sections of C$_2^-$ for Model~4 with different
values of ($\alpha_0$,$\beta$).}
\label{fig:mod4}
\end{center}
\end{figure}

\Fref{fig:2B3g} gives a series of Model 4 calculations for
$^2$B$_{3g}$ symmetry contribution to the total electron impact
electron detachment cross section as a function of energy. Both
singlet and triplet contributions are dominated by a series of
resonance features: several narrow resonances which show a strong
dependence on the PCO basis set parameters and a single broad
resonance in each case. In contrast to the narrow resonances, the
position and width of the broad resonance features is stable with
respect to the choice of PCO basis; indeed even the calculation with
$\alpha_0$=0.17 and $\beta$=1.5 which shows a much enhanced resonance
structure, a known signature of a poorly converged PCO basis
\cite{bhs96}, gives a similar resonance to the other calculations.
\Fref{fig:mod4} gives a similar comparison for the total
electron impact detachment cross section where again both the
singlet and triplet resonance features are clearly visible. The
poorer quality of the calculation based on the use of C$_2$ NOs can
clearly be seen.

\begin{table}
\caption{Energy, $E_r$, and width, $\Gamma$, both in eV, of resonance states of
C$_2^{2-}$; our results are based calculations with $\alpha_0$=0.17 and
$\beta$=1.4.}
\label{tab:res}
\begin{indented}
\lineup
\item[]\begin{tabular}{crccccrrrrrr}
\br
\ms
\centre{4}{Previous work}&& &  \centre{2}{Model 4} &\centre{2}{Model 5} & \centre{2}{Model 6}\\ 
\bs
\crule{4}&&\crule{7}\\
Symmetry&$E_r$ & $\Gamma$&ref&&Symmetry&$E_r$& $\Gamma$&$E_r$& $\Gamma$&$E_r$& $\Gamma$\\
\mr
$^1\Sigma_g^+$&3.5&0.3&$a$&&$^1\Sigma_g^+$&4.86&0.65& 4.75& 0.57&4.91 &0.67 \\
             &10.0&2.1&$b$&&$^1\Pi_g$&10.92&0.52&10.59&0.57&11.46&0.82 \\
             &10.0&3 -- 4&$b$&&$^3\Pi_g$&9.71&1.14&9.54&0.99&10.44&1.43\\
\br
\end{tabular}\\
\item[] $^{\rm a}$ Theory \cite{stm00}.
\item[] $^{\rm b}$ Experiment \cite{ahk96,pdj98,pdj99}.
\end{indented}
\end{table}

Inspection of plots of eigenphase sums (not given), showed one further
clear resonance at energies just above the ionisation threshold of
$^1\Sigma_g^+$ symmetry. Parameters for this resonance and the ones near
10~eV were obtained by fitting their eigenphases to a Breit-Wigner formula
\cite{jt31}. The resulting resonance parameters are given for Model 4
in \tref{tab:res}, where they are compared to previous work.

\citeasnoun{stm00} performed careful absorbing potential calculations
which found a single $^1\Sigma_g^+$ resonance close the ionisation
threshold of C$_2^-$.  We predict a similar resonance, albeit at
slightly higher energy; this agreement can be considered satisfactory
given the lower level of configuration interaction included in our
study. However we find two further resonances not found by
\citeasnoun{stm00}. These resonances lie close in energy to those
detected experimentally \cite{ahk96,pdj98,pdj99}. In fact the experiments
give somewhat different resonance characteristics depending on whether
they measure the detachment cross section, which gives a width of 2.1~eV,
or the smaller dissociation cross section, which gives widths between
3 and 4~eV. However to really test whether these resonances agree it
is necessary to make a more direct comparison with the experiments which
means comparing with the measured cross sections.

\subsection{Cross sections}

\begin{figure}
\begin{center}
{\resizebox{100mm}{!}{ \includegraphics*{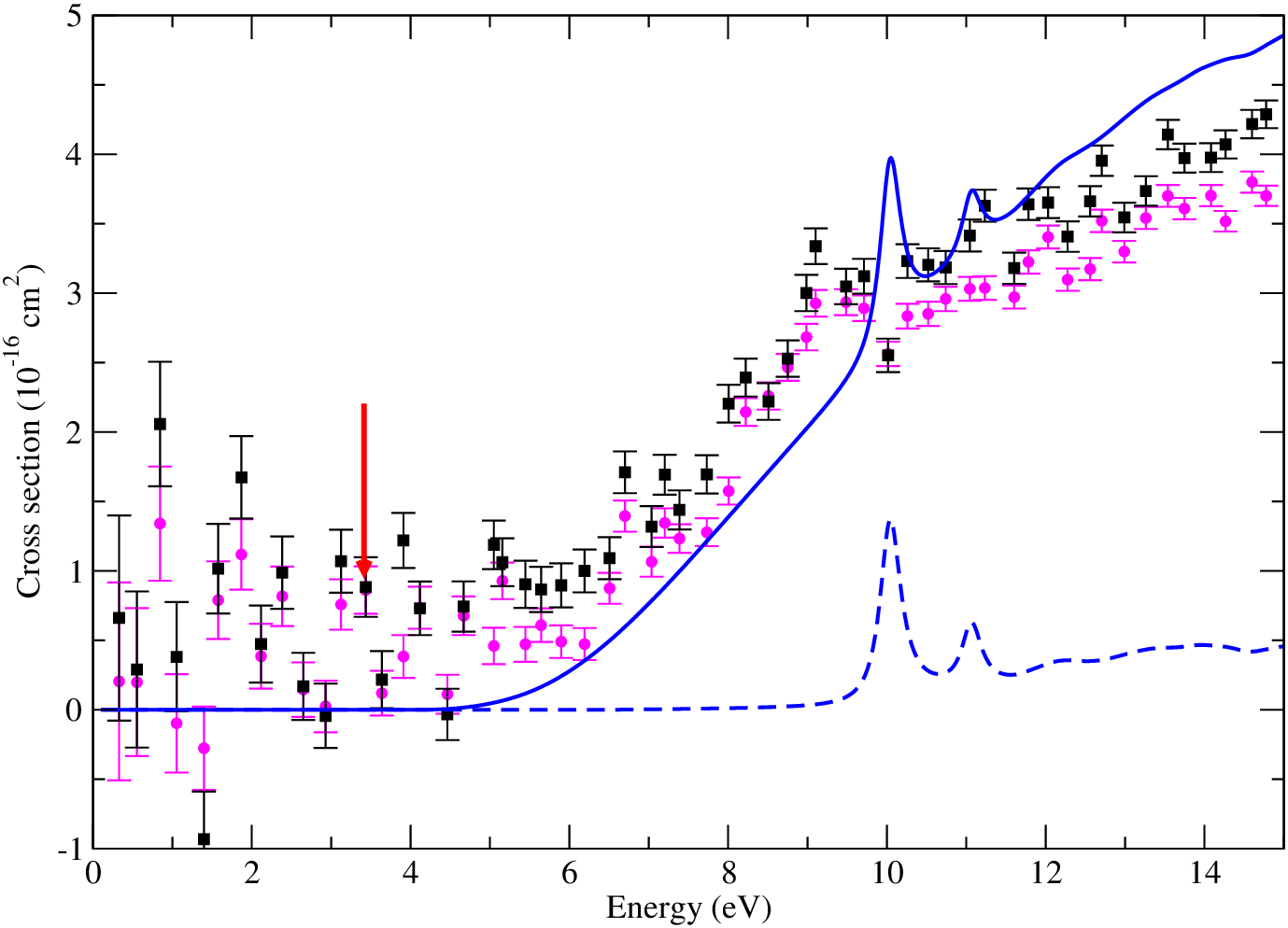}}}
{\resizebox{100mm}{!}{ \includegraphics*{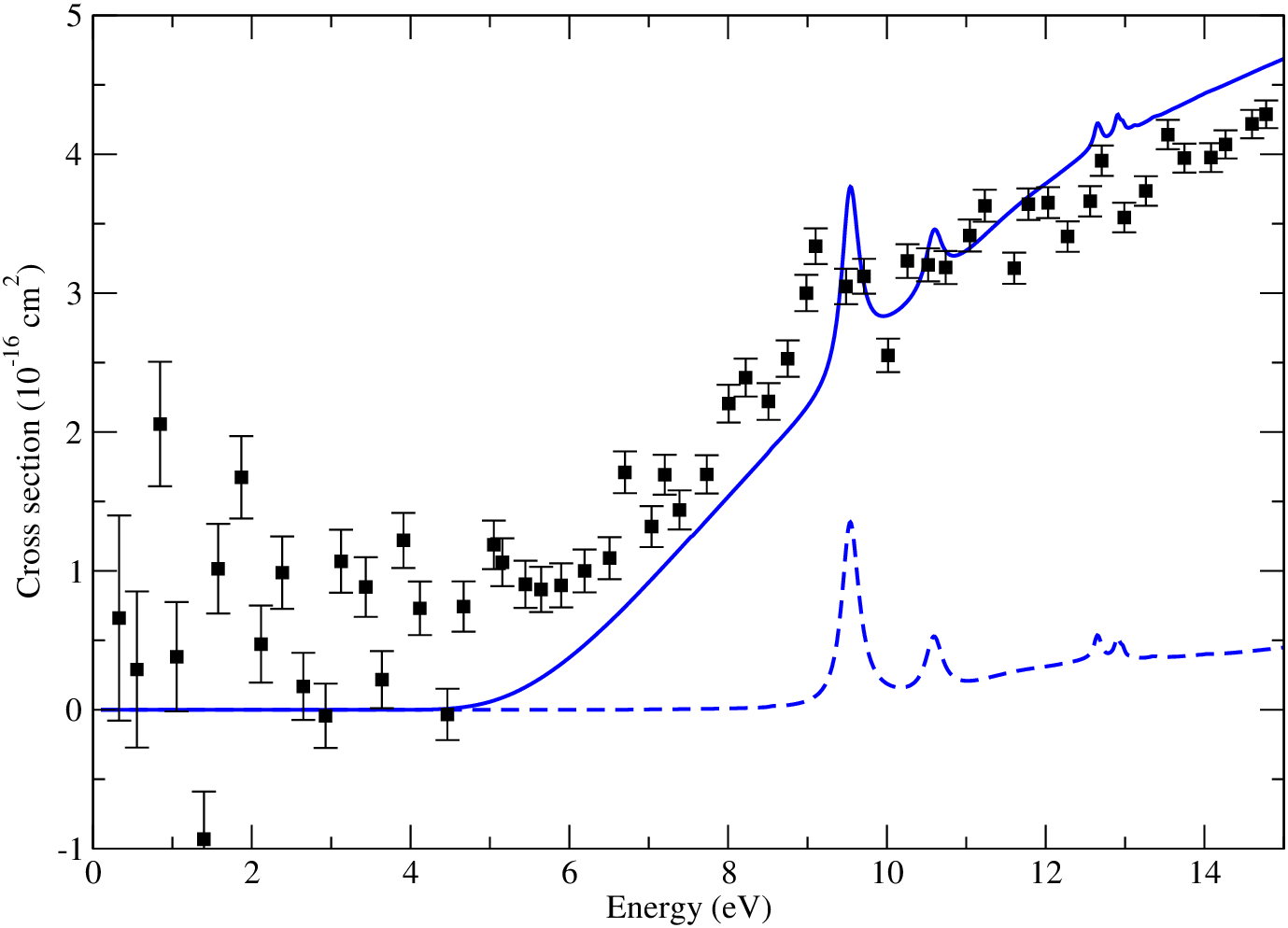}}}
{\resizebox{100mm}{!}{ \includegraphics*{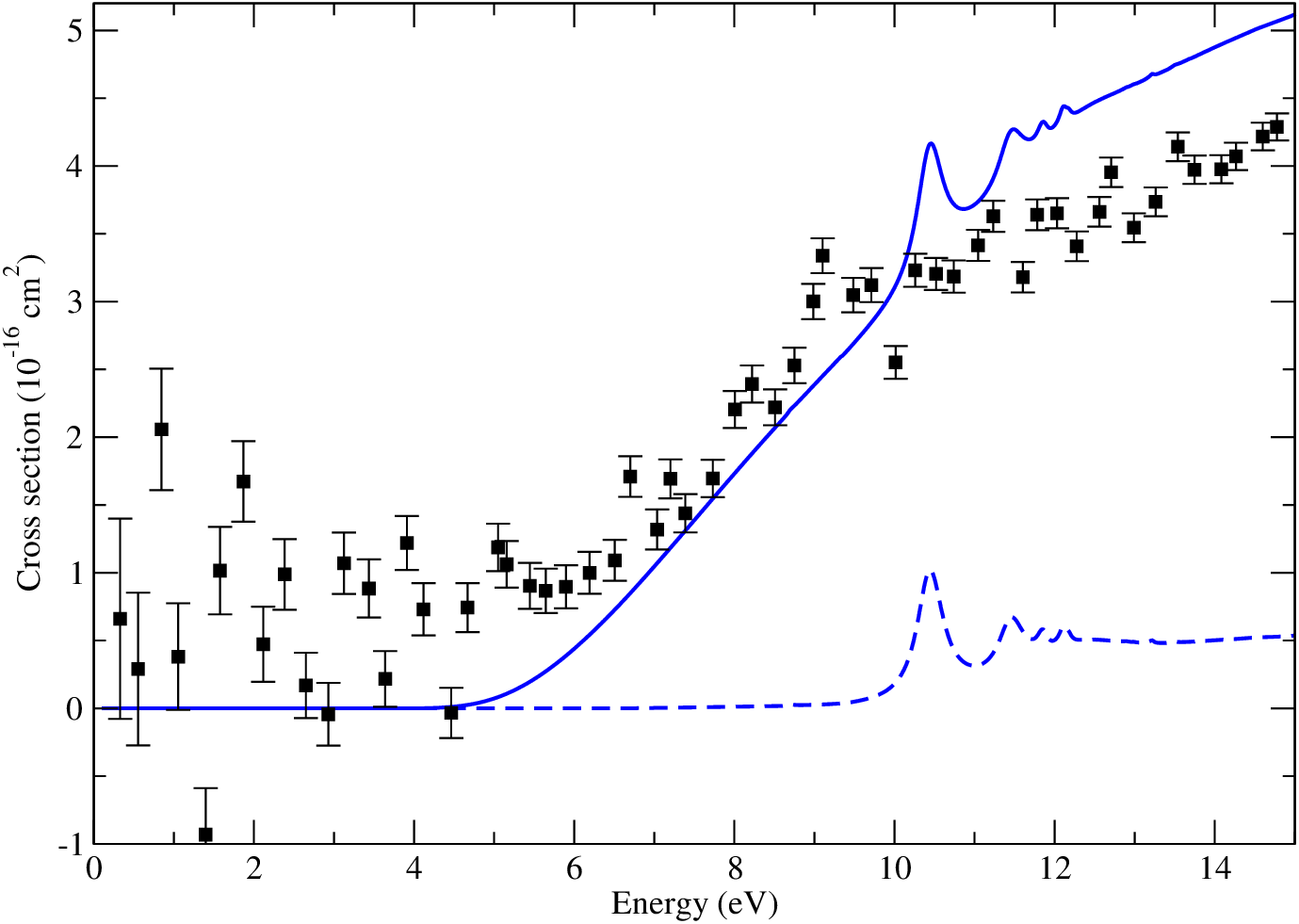}}}
\caption{C$_2^-$ electron impact ionisation cross sections. 
Experimental data due to \citeasnoun{pdj99} is given by circles
(without dissociative channels) and squares (with dissociative
channels); the dashed lines represent our cross sections without Born
correction; the solid line represent our cross sections with Born
correction.  The arrow indicates the location of the ionisation
threshold.  Top panel: Model~4 with $\alpha_0$=0.17 and $\beta$=1.4;
Middle panel: Model~5 with $\alpha_0$=0.17 and $\beta$=1.4; Bottom
panel: Model~6 with $\alpha_0$=0.17 and $\beta$=1.3.}
\label{fig:totion}
\end{center}
\end{figure}

\Fref{fig:totion} gives a comparison between our calculations, performed
with three different models, and the measurements. What is immediately
apparent is that without the Born correction our cross sections are very
significantly lower than the measured ones. This supports the assertion
made above that the majority of the ionisation occurs through large impact
parameter collisions.

All the models shown give reasonable agreement with the measurements
of \citeasnoun{pdj99}. For completeness we give the measured results
with and without the consideration of simultaneous ionisation and dissociation.
Our calculations implicitly give the cross section for all processes that 
involve ionisation and therefore should account for both channels.
As the dissociation channel is relatively unfavoured, these two
measurements are close together, the difference is certainly smaller than
the accuracy of our calculations.

It can be seen that all three models give total ionisation cross sections
in good agreement with the measurements. In this context it should be noted
that the cross sections should all go to zero below the ionisation threshold
denoted by the arrow. Furthermore our resonance structures are very
similar to, if a little higher in energy than, the observed ones. For
reasons discussed for the lower-lying $^1\Sigma_g^+$ resonance, we would
expect our resonance energies to be a little too high. This aspect
of the agreement can therefore also be considered to be very good.

\begin{figure}
\begin{center}
{\resizebox{130mm}{!}{ \includegraphics*{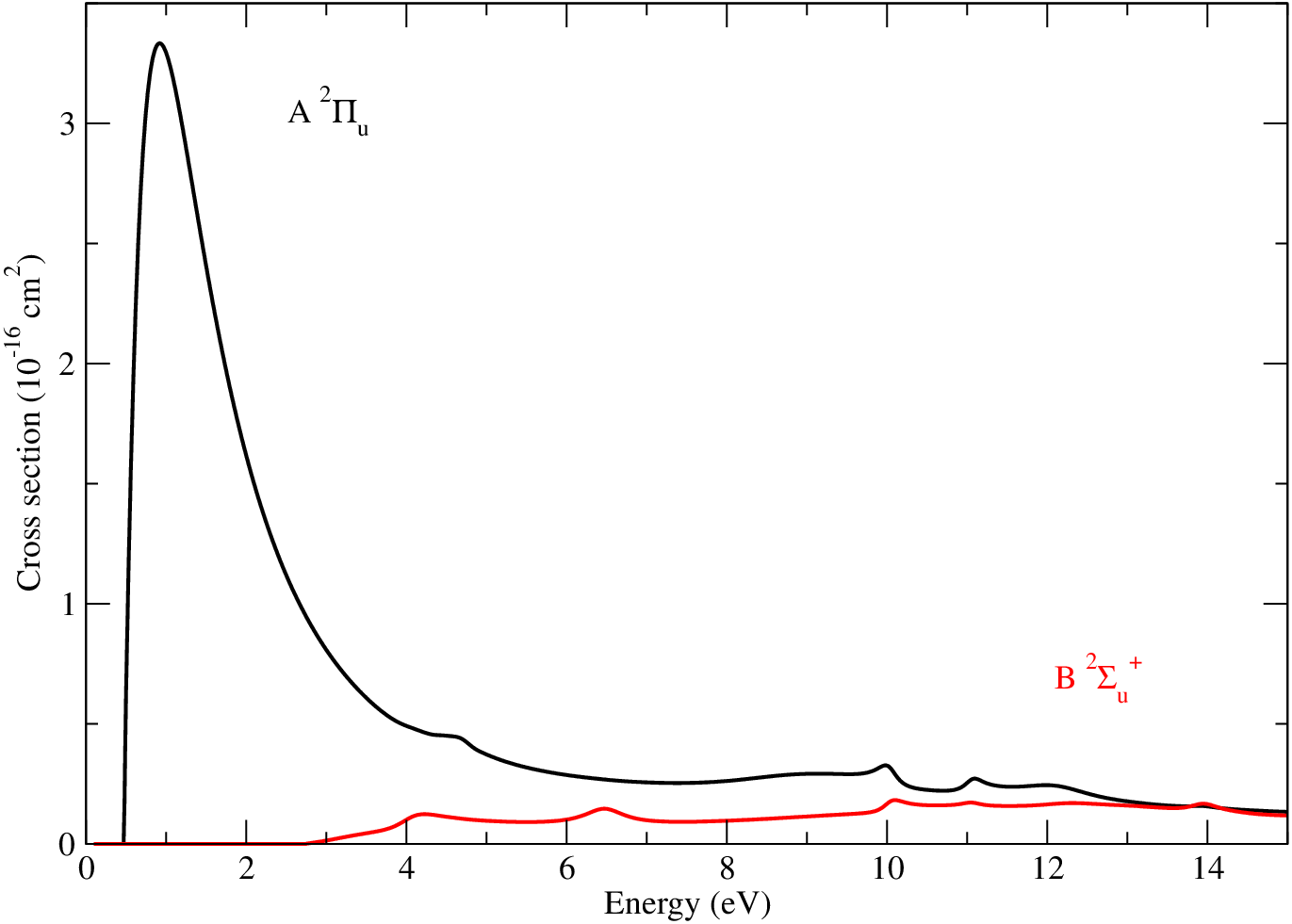}}}
{\resizebox{130mm}{!}{ \includegraphics*{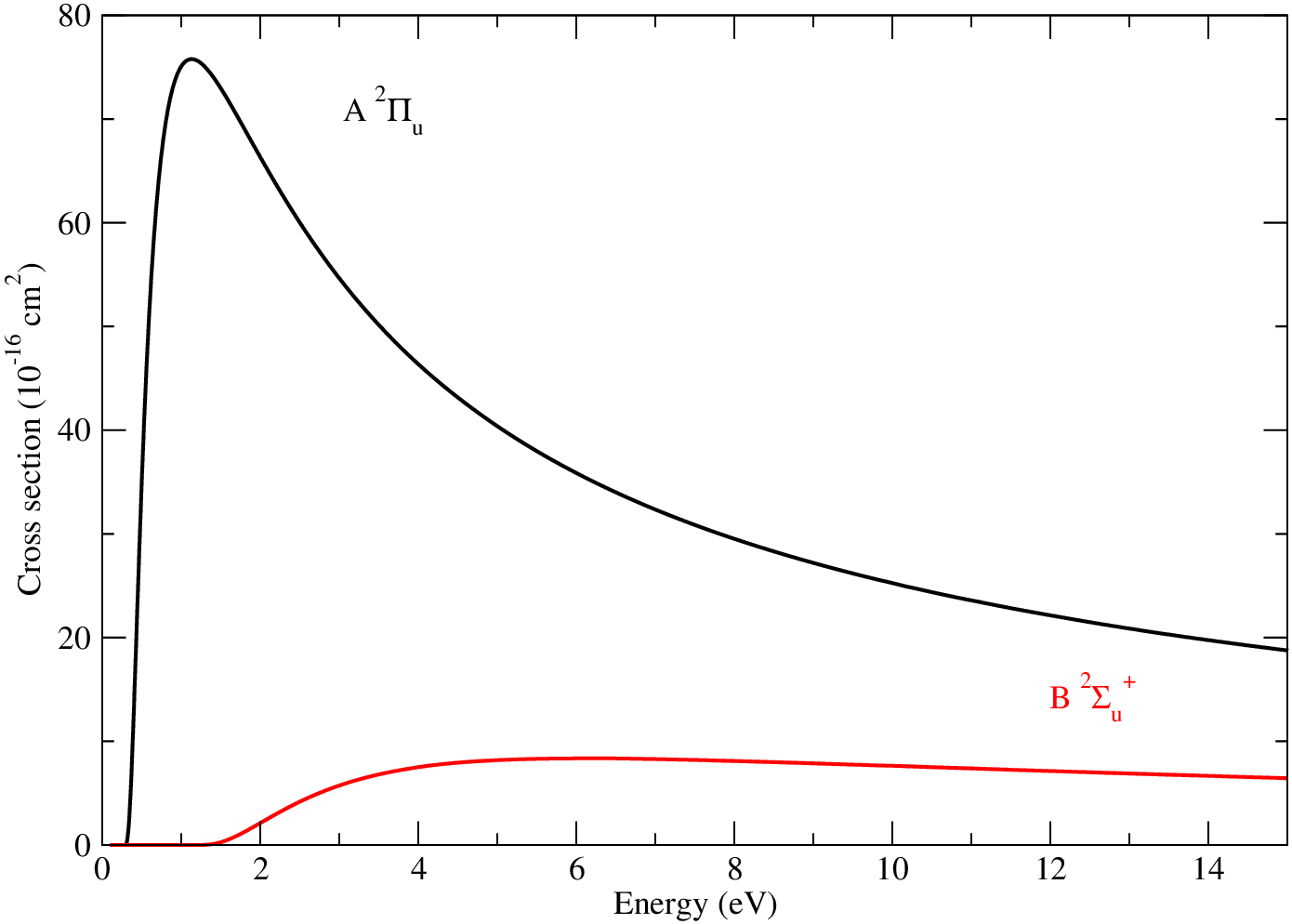}}}
\caption{Electron impact electronic excitation cross sections for excitation
to the $A$~$^2\Pi_u$ and $B$~$^2\Sigma_u^+$ states without (upper) and
with (lower) Born correction; note the very different cross section
scales for the two panels. Calculations are for Model~4 with
$\alpha_0$=0.17 and $\beta$=1.3.}
\label{fig:toex}
\end{center}
\end{figure}

Finally \fref{fig:toex} gives total electron
impact electronic excitation cross sections for the two bound states of
C$_2^-$. These excitations are both dipole allowed and the cross sections
are again completely dominated by the high impact parameter, dipole-driven
transitions which we characterise with the Born approximation. Indeed
as we predict essentially no resonance structure in these cross sections,
a simple Born calculation should suffice for this problem.

\section{Discussion and conclusions}

Our calculations detect the clear signature of 3 resonances: one of
$^1\Sigma^+_g$ symmetry just above the ionisation threshold and
resonances with singlet and triplet spin and $\Pi_g$ spatial symmetry
at about 10~eV.  Since they all lie above the excited electron states
of C$_2^-$, these must be regarded as shape resonances. The standard
model for shape resonances in electron -- molecule collisions relies
on a centrifugal barrier to provide the trapping potential. Such
resonances have two characteristics: they are not found for s-wave
scattering and they occur, often at too high an energy, in simple
static exchange calculations. The present resonances are absent from
the test static exchange calculations we ran and the $^1\Sigma^+_g$
resonance involves s-wave scattering.

All this implies a modified picture of a shape resonance. The situation
here is that the scattering occurs under the influence of a strongly
repulsive Coulomb potential. The well required to (temporarily)
trap the scattering electron should therefore be the result of some local
attraction. Our calculations suggest that this is indeed what happens
and the attractive component is provided by polarisation interactions.
This is why characterising the resonances require such a sophisticated
representation of the target wave functions. 

Although representation of the resonances requires a very sophisticated
treatment of the target, this is not the situation for the cross sections.
Both the electron impact electronic excitation and electron detachment
cross sections are dominated by long range collisions which can be modelled
using a simple Born approximation. It should however be noted that 
our treatment of the electron impact electron detachment process does involve
the use of a discretised continuum via the MRMPS method which is then used
to provide the necessary transition dipoles for the Born calculation.

In summary our calculations on electron impact electron detachment of C$_2^-$
reproduce both the resonance structure and cross sections observed over
a number of years in storage rings  \cite{ahk96,pdj98,pdj99} for this
process. These calculations are the first to achieve either of these goals. To
do this we had to combine a sophisticated treatment of the C$_2^-$
target including representation of target continuum states using the
molecular R-matrix with pseudo states (MRMPS) method with a method for
calculating the effects of high impact parameter collisions.

\begin{section}*{References}
%\bibliography{c22m,jt,c2,rmps,rmat}

\end{section}

\end{document}